\documentclass[preprint]{sigplanconf}
\usepackage{graphicx}
\usepackage{eepic}

\begin{document}

\title{Actor Continuation Passing}
\subtitle{Efficient and Extensible Request Routing for Event-Driven Architectures}

\authorinfo{Stefan Plantikow}{Zuse Institute Berlin (ZIB)}{stefan.plantikow@googlemail.com}

\date{16 January 2010}

\maketitle

\begin{abstract} The logic for handling of application requests to a staged, event-driven
architecture is often distributed over different portions of the source code. This complicates
changing and understanding the flow of events in the system.

The article presents an approach that extracts request handling logic from regular stage
functionality into a set of request scripts. These scripts are executed step-wise by sending
continuations that encapsulate their request's current execution state to stages for local
processing and optional forwarding of follow-up continuations. A new domain specific language that
aims to simplify writing of request scripts is described along with its implementation for the scala
actors library. Evaluation results indicate that request handling with actor continuations performs
about equally or better compared to using separate stages for request handling logic for scripts of
at least 3 sequential steps.
\end{abstract}

\category{H.2.4}{Information Systems}{Systems}[Concurrency]         
\category{D.1.3}{Software}{Programming Techniques}[Concurrent Programming]         
\category{D.3.3}{Programming Languages}{Language Constructs and Features}[Concurrency]

\keywords{Request Routing, Staged Event-Driven Architecture, Continuations, Actor Model, Scala}

\section{Introduction}             

Staged, event-driven architectures~\cite{Welsh2001} implement an approach to the design of server
software that can provide high degrees of concurrency and throughput. This is achieved by
structuring the software as a set of stage that run in separate threads, do not share state, and
communicate exclusively via event queues, i.e. follow the actor model of message passing
concurrency.

Requests to the server application are enqueued as events at some stage. Handling such an event may
involve pure local computation, accessing stage-specific functionality (manipulation of local state
and resources), and continuing request handling by sending new events to other stages. This
\emph{application logic} can be divided into \emph{stage logic} which must necessarily be executed
at a fixed stage and \emph{request logic} which may be executed anywhere as long as it is provided
with the required inputs.

The resulting interactions during request handling at runtime can be complex and difficult to
understand. Therefore it appears desirable that at least all request logic for a given request type
should be implemented in a readable, singular section of the source code. However application logic
is typically spread over the implementations of all stages.  Additionally, adding new request types 
may require the introduction of new event types to communicate intermediary values between stages.

This distribution of application logic over different stages reduces the understandability of staged
architectures and complicates modifying the handling of application requests. Additionally, it
impedes the addition of new request types without changing the source code of existing stages and
redeploying parts of the system.

In this article, an approach for extracting request logic into separate source code units that are
independent of the implementation of stage logic is presented. The solution is based on sending
continuations between stages and CPS-transformation of request handling code. It is unique
in that it does neither require additional messages nor leads to source code with deeply nested
callbacks. The approach has been implemented as a domain specific languages~(DSL) for the scala
actors library~\cite{Haller2007}.
                                                 
%
%
%
                                
\section{Intertwined Logic}

The distribution of application logic over the system stems from the intertwining of
stage-independent global \emph{request logic} and local \emph{stage logic} that actually has to be
performed at a specific stage. For a given request, each part of request logic is glued together
with some stage logic quite randomly as chosen by the developer. Required intermediary values are
sent as part of the event that triggers a block's execution.

Additionally, the result of executing stage logic may determine how and at which concrete stages
request handling needs to be continued. This places further burdens on the implementation of
intermediary stages, as incoming- and outgoing event types have to be amended with request state,
although it might be completely independent from the intended purpose of that stage.

To give an example for this, imagine a simplified system for launching satellites into space.
Incoming reqeuests are amended with authentication information in the first stage. In the second
stage, this information is then used to authorize the request and eventually launch the rocket. Only
after the satellite has begun to operate, some third stage (i.e. the press office) is informed.

First, note how the launching stage needs to know about the overall workflow in order to forward a
message to the press office after a succesful launch. Now, imagine that the initial request needs to
be amended with extra information (name and owner of satellite) for the press office. Passing this
information down requires modifying the events to and from the rocket launching stage with fields
for the additional payload, although this extra information is of no importance to actually
launching the rocket.

This intertwining of request and stage logic is a case of insufficient separation of concerns,
calling for a different way to describe both types of application logic.

\section{Separating Request and Stage}

Extracting request logic requires a mechanism for interruptible, stateful control-flow. One approach
to this is the use of additional coordination stages. In this scheme, request logic is executed by a
coordinator stage. Stage logic is executed by sending and receiving events to the executing stage.
This solution has some drawbacks: Executing stage logic requires the sending of two messages
(request-response), leads to an additional thread context switch back to the coordinator stage (e.g.
to access pre-request state), and may necessitate the introduction of new event types to communicate
intermediary values. Additionally, special care may be required to avoid overloading of coordinator
stages by implementing them in a non-blocking fashion and load-balancing over them.

The \emph{continuation} of a computation at a point in time describes the part of a computation that
yet needs to be computed, i.e. describes stack frame state and remaining program. A continuation may
be explicitly stored as a value by reifying it as an anonymous function. It may then be restored and
executed arbitrary often by calling this function.

This allows to implement pausable, stateful control-flow: Stages receive events that actually are
anonymous functions that represent the current continuation of some request. Such incoming
continuations are executed by calling them with the executing stage as their sole argument. Thus
request continuations gain access to the functionality of local stages. When the execution of a
request continuation is about to finish, optionally, the follow-up continuation may be captured in a
last step and sent to the next stage where request handling continues.

This \emph{actor continuation passing (ACP)} approach does not require any intermediate stages for
the execution of the request logic and thus avoids the introduction of additional messages. It does
not require special events for communicating intermediary values since they are contained in the
continuation stack frames. On the downside, it requires some overhead for continuation capturing.

\section{A Request Handling DSL}

Next, Souffleuse, a library for request handling with actor continuation passing is presented.
Souffleuse has been implemented in the scala programming language~\cite{Odersky2004} using the
scala actor library~\cite{Haller2007}. Souffleuse provides a \emph{domain specific language (DSL)}
for writing request scripts that execute over a set of locally running stages. Request scripts are
implemented in terms of a simple set of commands that allow structuring scripts as a sequence of
code blocks that are executed at different stages.

\begin{itemize}                                   
	\item \textbf{\emph{v} $\mathbf{\leftarrow}$ remember(\emph{value})} Bind \emph{value} to \emph{v} for later reuse by
	the script
	\item \textbf{\emph{v} $\mathbf{\leftarrow}$ compute(\emph{thunk})} Compute \emph{thunk} at the current local stage.  
	The return value is bound to \emph{v} and may be reused later in the script
	\item \textbf{\emph{s} $\mathbf{\leftarrow}$ goto(\emph{stage})} Continue  request execution at stage \emph{stage} and 
	return a reference \emph{s} for gainining access to stage functionality of \emph{stage} 
	(usually \emph{stage} itself)
	\item \textbf{yield(result)} The yield statement of the \textbf{for}-expression may optionally be used to 
	return a result to the initial caller of the script                           	
\end{itemize}

Routing scripts may be written as \textbf{for}-expressions and are executed using two additional
primitives of the DSL:

\begin{itemize}
    \item \textbf{run(\emph{forExpr})} Run \emph{forExpr} and wait until its execution yields a result (blocks current actor)
	\item \textbf{asyncRun(\emph{forExpr})} Runs \emph{forExpr} without waiting for a result (non-blocking)
\end{itemize}	

As an example, consider the execution of a single remote procedure call: 

\medskip {\footnotesize\begin{tabular}{l} $\textbf{def}\ $rpc$(\emph{targetStage}, \emph{args})\ =\
\{$\\ \hspace{2ex} $\textbf{val}\ \emph{request}\ =\ \textbf{for} ($\\ \hspace{6ex}
$\emph{stageRep}\ \leftarrow\ $\textbf{goto}$(\emph{targetStage})$\\ \hspace{6ex}
$\emph{procResult. }\ \leftarrow\ $\textbf{compute}$\ \{\ stageRep.$proc$(\emph{args})\ \}$\\
\hspace{2ex} $)\ \textbf{yield}\ procResult$\\ \hspace{2ex} $\textbf{return}\ $run$(\emph{request})\
\}$\\ \end{tabular}} \medskip

The call is wrapped as a regular function that initially assembles a new request script. The script
itself first transfers the execution to the \emph{targetStage} for the RPC using \textbf{goto}.
Then, the actual RPC is executed at that stage using \textbf{compute}, and finally the return value
is yielded. To actually execute this request script, it is started with \textbf{run}.

Alternatively, request scripts may be written by subclassing the class Play and overriding its
\textbf{apply} method. This allows to place an upper type bound on all stage instances used by the
script.

Stages are implemented by subclassing or instantiating the class Stage (a stock scala actor) and
providing it with an exchangeable stage functionality object (called its \emph{Prop}) that is passed
to each request script executing at that stage. The Prop instance may be identical with the Stage
itself.

\section{Implementation}                                   

Soeuffleuse implements request handling according to the actor continuation passing approach on top
of the scala actor library. Stages are implemented as actors that run in separate
threads.\footnote{Stages are receive-actors in terms of the scala actor library} Their main loop
listens for messages consisting of one-argument anonymous lambda functions. When such a function is
received, it is executed and given access to the stage by passing the prop as its first argument.
However, explicitely writing out continuation functions can lead to unreadable source code with a
nesting level of anonymous lambda functions that is as large as the number of sequentially passed
stages.

As a remedy, Souffleuse performs CPS-transform and sending of continuations at stage boundaries.
\emph{Continuation Passing Style (CPS)} refers to a control flow graph transformation that replaces
regular function return with calling a continuation function passed as an extra argument. Scala's
\textbf{for}-generator-expressions provide generator objects with continuation functions for the
remainer of the for-loop through a CPS-transform done by the scala compiler. How these continuatios
are called is left to the generator. This is exploited by Souffleuse's \textbf{goto} command to
capture the current continuation and send it to a remote stage for execution.    
        
\begin{figure*}[t]    
\begin{center}
\includegraphics[width=.33\hsize]{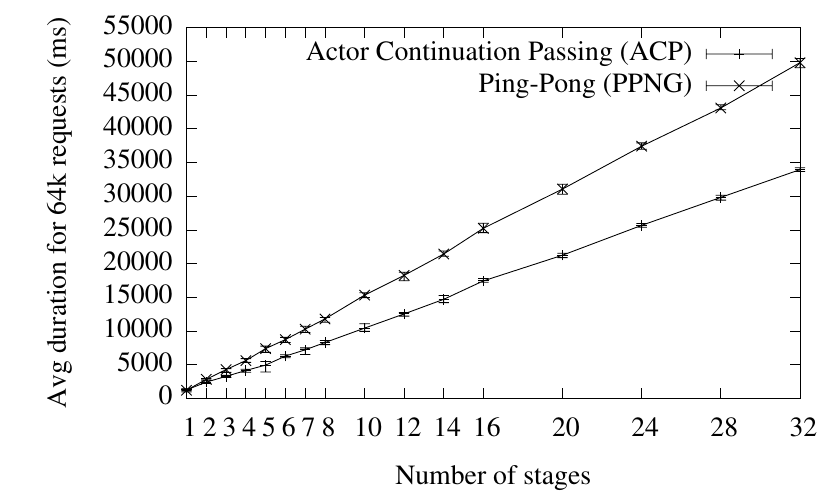}%
\includegraphics[width=.33\hsize]{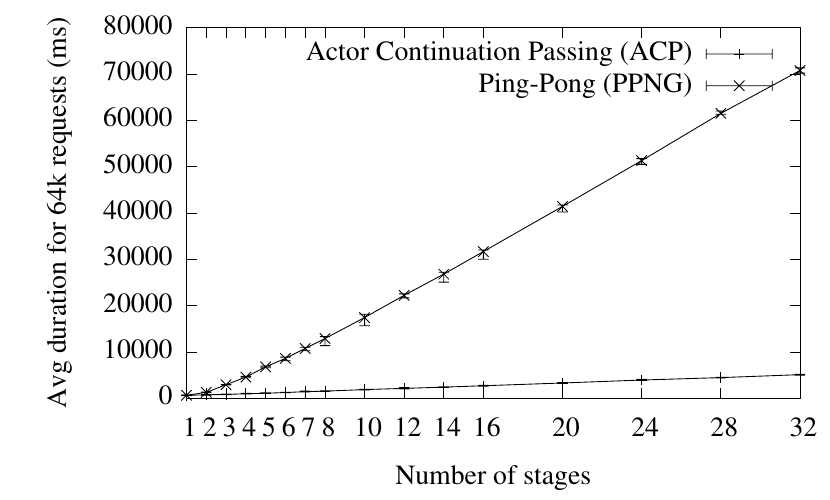}%
\includegraphics[width=.33\hsize]{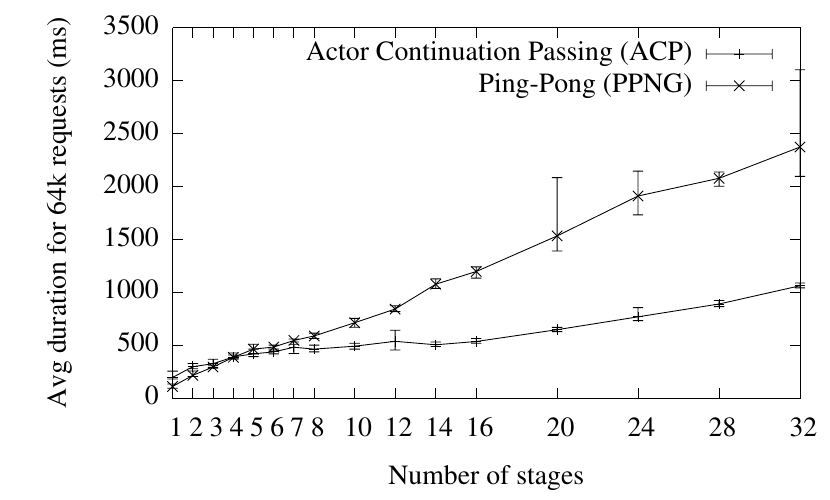}
\end{center}
\caption{Comparing Actor Continuation Passing against using a coordinator stage when sending messages around a ring\label{fig:eval}} 
\end{figure*}

\subsection{CPS-Transform in Scala}     

Next it is shown how Souffleuse exploits Scala's \textbf{for}-generator-expressions to implement
CPS-transform. Routing scripts are written as expressions of the form:

\medskip
{\footnotesize\begin{tabular}{l} $\mathbf{for}\ (\mathit{v_1} \leftarrow \mathit{e_1}, \mathit{v_2} \leftarrow
\mathit{e_2}, \ldots, \mathit{v_n} \leftarrow \mathit{e_n})\ \mathbf{yield}\ \mathit{r}$
\end{tabular}}
\medskip

This iterates sequentially from outmost to innermost over the generators $e_i$. Each $v_i$ is
consecutively bound to the values produced by its generator $e_i$. Results are created by evaluating
$r$ in each iteration until $e_1$ is exhausted.

Scala abstracts from how \textbf{for} interprets different types of generators by CPS-transforming
the expression and calling abstract methods on the generators. For example, above expression is
transformed by the Scala compiler into:

\medskip
{\footnotesize\begin{tabular}{l}
$\mathit{e_1}.$flatMap$\ \{\ \mathbf{case}\ \mathit{v_1}\ \Rightarrow$\\
$\hspace{2ex}\mathit{e_2}.$flatMap$\ \{\ \mathbf{case}\ \mathit{v_2}\ \Rightarrow\ \ldots\ \mathit{e_n}.$map$\ \{\ \mathit{r}\ \}\ \ldots\ \}\ \}$
\end{tabular}}
\medskip

Every $\{\ \mathbf{case}\ \mathit{v_i}\ \Rightarrow\ \ldots\ \}$ is an anonymous lambda function
that reifies the continuation for the remaining \textbf{for}-generator-expression. To make this
implicit CPS-transformation usable, the scala standard library contains the abstract class Responder
which provides implementations of flatMap and map in terms of a function respond. Respond takes the
continuation for the remaining generator-expression as its only argument, generates values, and
iterates by calling the continuation with them.

To implement actor continuation passing, Souffleuse associates each stage with a Responder whose
respond method simply forwards passed continuations to the stage via message passing:
                                                                                            
\medskip
{\footnotesize\begin{tabular}{l}                                
$\mathbf{object}\ $responder$\ \mathbf{extends}\ $Responder[this.type]$\ \{$\\
\hspace{2ex}$\mathbf{def}\ $respond$(\mathit{k}:\ $Actor.this.type$\ \Rightarrow\ $Unit$):\ $Unit$\ =\ $self.send$(\mathit{k})$\\
$\}$\\
\\
$\mathbf{def}\ $asResponder$:\ $Responder[Actor.this.type]$\ =\ $responder\\
\end{tabular}}
\medskip

This mechanism is sufficient to implement the Souffleuse DSL. \textbf{goto} returns a responder for
its argument as described above. \textbf{remember} simply creates a constant responder for a single
value. \textbf{run} uses the actor library to create a dedicated channel for return values. All
other commands are implementable on top of goto and remember.


\subsection{Continuation Access}

Souffleuse features a type of Stage that allows routing scripts to access the currently running
continuation. This may be used to forward continuations to other stages (load balancing), execute
the same continuation repeatedly over multiple actors (replication).

\subsection{Limitations and future work}

The strictly linear nature of generator expressions makes writing routing scripts with
non-linear control-flow more difficult and may require the execution of routing sub-scripts.  
However, even in such a scenario all request routing logic is written in a single routing
script.

Souffleuse currently does not yet support exception handling across stage boundaries since the
correct specification of such a facility is not obvious to the author at this point, especially
considering non-linear request scripts with synchronization.

Beyond exception handling, it would be desirable to extend the library for synchronization in the
case of non-linear request scripts. This raises interesting questions in terms of garbage collection
and management of auxiliary state by stages that are used as synchronization points.

\subsection{Availability}

Souffleuse is being made available as open source.

\section{Evaluation}
     
Souffleuse has been compared against using coordinator stages in a
messages-around-a-ring-of-stages-scenario with different load generation strategies. The results
indicate that it performs equally or better when the ring size is $\ge$~3~(Fig.~\ref{fig:eval}).
With growing ring size, Soufflese converges towards a twofold performance increase since the
required number of messages is halved. The use of events consisting of serialized continuation
functions appears to have neglible overhead. All tests were conducted on a 8-core machine.

\section{Conclusion}

The results indicate that actor continuation passing is a viable approach for separating request
from stage logic without suffering from the performance penalty introduced by using explicit
coordinator stages. Souffleuse implements this approach as a mini-DSL in scala. The implementation
eliminates the need for deeply nested callbacks by using the implicit CPS-transformation of scala's
\textbf{for}-expression. Thus writing request logic as fast and concise request scripts is enabled,
while stage logic is implemented separately where it belongs.

\subsubsection*{Acknowledgments}

The author thanks Bj\"orn Kolbeck who initially described the problem in the context of
XtreemFS~\cite{Hupfeld2007}, a distributed filesystem implemented as a staged architecture.

\bibliographystyle{abbrv} 
\bibliography{sigproc}

\begin{thebibliography}{1}

\bibitem{Haller2007}
P.~Haller and M.~Odersky.
\newblock Actors that unify threads and events.
\newblock {\em LNCS}, 4467, Jan 2007.

\bibitem{Hupfeld2007}
F.~Hupfeld, T.~Cortes, B.~Kolbeck, E.~Focht, M.~Hess, J.~Malo, J.~Marti,
  J.~Stender, and E.~Cesario.
\newblock Xtreemfs - a case for object-based storage in grid data management.
\newblock In {\em Proceedings of 33th International Conference on Very Large
  Data Bases (VLDB) Workshops}, 2007.

\bibitem{Odersky2004}
M.~Odersky, P.~Altherr, V.~Cremet, B.~Emir, and S.~Maneth.
\newblock An overview of the scala programming language.
\newblock Technical Report LAMP-EPFL 2006-001, EPFL, Jan 2006.

\bibitem{Welsh2001}
M.~Welsh, D.~Culler, and E.~Brewer.
\newblock Seda: An architecture for well-conditioned, scalable internet
  services.
\newblock In {\em Proceedings of SOSP 18}, 2001.

\end{thebibliography}

\end{document}